\documentclass[aps,prd,preprint,numerical,showpacs,showkeys,superscriptaddress]{revtex4-1}

\usepackage{amsmath}
\usepackage{amssymb}
\usepackage{amsfonts}
\usepackage{graphicx}
\usepackage{dcolumn}
\usepackage{bm}
\usepackage{hyperref}
\usepackage{wasysym}
\hypersetup{colorlinks=false}

\begin{document}

\title[Spacetime geometry fluctuations and geodesic deviation]{Spacetime geometry fluctuations and geodesic deviation}

\date{\today}

\author{H. S. Vieira}
\email{Horacio.Vieira@tufts.edu and horacio.santana.vieira@hotmail.com}
\affiliation{Institute of Cosmology, Department of Physics and Astronomy, Tufts University, Medford, Massachusetts 02155, USA}
\affiliation{Departamento de F\'{i}sica, Universidade Federal da Para\'{i}ba, Caixa Postal 5008, CEP 58051-970, Jo\~{a}o Pessoa, PB, Brazil}
\author{L. H. Ford}
\email{ford@cosmos.phy.tufts.edu}
\affiliation{Institute of Cosmology, Department of Physics and Astronomy, Tufts University, Medford, Massachusetts 02155, USA}
\author{V. B. Bezerra}
\email{valdir@fisica.ufpb.br}
\affiliation{Departamento de F\'{i}sica, Universidade Federal da Para\'{i}ba, Caixa Postal 5008, CEP 58051-970, Jo\~{a}o Pessoa, PB, Brazil}

\begin{abstract}
The quantum fluctuations of the geodesic deviation equation in a flat background spacetime are discussed. We calculate the resulting mean squared fluctuations
 in the relative velocity and separation of test particles. The effect of these quantum fluctuations of the spacetime geometry is given in terms of the 
 Riemann tensor correlation function. Three different sources of the Riemann tensor fluctuations are considered:  a thermal bath of gravitons,  
 gravitons in a squeezed state, and the graviton vacuum state.
\end{abstract}




\maketitle


%
%
\section{Introduction}
In flat space, parallel lines maintain their separation forever. However, in curved spacetime, parallel geodesics do not remain parallel when are extended. 
The mathematical statement of this physical phenomena is given by the geodesic deviation equation, which shows that the tidal force of a gravitational field 
causes modification in the trajectories of neighboring particles \cite{Carmeli:1982}. Many studies concerning the behavior of the geodesic deviation equation 
in several background gravitational fields as well as their consequences can be found in 
\cite{PhysRevD.96.064013,EurPhysJC.77.372,PhysRevD.96.084020,ClassQuantumGrav.34.215003,PhysRevD.96.105004}.
In the general relativity theory, an important effect played by curvature is how it changes the relative separation between two geodesic particles. 
This is a manifestation of the gravitational field and hence the acceleration of the deviation vector between two nearby geodesics contains information 
about the curvature of the spacetime \cite{Carroll:2004,AstrophysJ.849.158,ClassQuantumGrav.34.215002,PhysRevD.96.083015}.

The properties of the curved spacetime which are reflected by physics in a gravitational field can be evaluated by analyzing the behavior of a set of neighboring geodesics, 
representing, for example, a bundle of photons or a distribution of massive test particles \cite{Padmanabhan:2010}. In order to study this phenomenon, many different 
approaches have been proposed \cite{PhysRevD.34.1014,PhysRevD.60.084018,ClassQuantumGrav.33.115002,ClassQuantumGrav.34.165003}.

On the other hand, the investigation of the Brownian motion, which can be described by the Langevin equation, played a very important role for the establishment of the 
atomic structure of matter. The discreteness character of matter (microscopic feature) causes fluctuations in the density of matter, which, in turn, causes observable effects 
on the motion of the Brownian particle (macroscopic feature) \cite{Kubo:1988}. Recently, the solutions of Langevin-type equations in some astrophysical scenarios have been 
discussed in the 
literature ~\cite{EurPhysJC.74.2900,PhysRevD.89.085037,EurPhysJC.76.160,PhysRevD.93.083507,MonNotRoyAstronSoc.457.3922,PhysRevD.93.043501,PhysRevD.95.103521}.

The knowledge of the behavior of a Brownian particle immersed in a fluid of much smaller atoms, can give us, in principle, some relevant information about the physics of these 
objects~\cite{Salinas:2001}. Brownian motion of test particles coupled to quantized fields was studied in Refs.~\cite{PhysRevD.70.065009,BBF08,PF11}.
Similarly, we can study the Brownian motion of test particles in a fluctuating gravitational field to look for insights into quantum gravity~\cite{PF14} .
In this way, we will use the geodesic deviation equation as a Langevin equation in which the Riemann tensor fluctuates. 
These quantum fluctuations of the curvature modify the motion of test particles and can be measured by the relative velocity dispersion after an interaction.

The quantum fluctuations of the spacetime geometry can be of two types: passive and active. The passive case is generated by fluctuations of the quantum matter fields, that is, 
from fluctuations in the source of the gravitational field which are described in terms of the stress and Ricci tensor correlation 
functions ~\cite{JRussLaserRes.26.445,PhysRevD.76.124018,PhysRevD.78.044025,PhysRevD.85.124014,PhysRevD.88.045011,PhysRevLett.111.060403,GenRelativGravit.47.24}. 
The active case is due to the quantum nature of gravity, that is, from fluctuations of the dynamical degrees of freedom of gravity itself, which are given in terms of the
 Riemann tensor correlation 
function~\cite{JPhysConfSer.174.012015,PhysRevLett.107.021303,FoundPhys.41.77,IntJModPhysD.21.1242012,PhysRevD.85.104048,IntJModPhysConfSer.12.299,PhysRevD.89.024039,ClassQuantumGrav.32.185019}.

This paper is organized as follows. In Section~\ref{sec:geodesic} we introduce the geodesic deviation equation and obtain an expression for the relative velocity dispersion. 
In Section~\ref{sec:thermal}, we evaluate this expression and compute the relative distance fluctuations in the case of a thermal bath of gravitons. In Section~\ref{sec:squeezed}, 
we do the same for gravitons in a squeezed state. In Section~\ref{sec:vacuum}, we sample the Riemann tensor correlation function for the case of the graviton vacuum state. 
Finally, in Section~\ref{sec:final}, the conclusions are given. We will use units in which $\hbar = c  = 1$ throughout the paper. In Sections~\ref{sec:thermal} and \ref{sec:vacuum},
we use units in which $32 \pi G = 32 \pi \ell_{\mbox{\tiny Pl}}^2 =1$ where $G$ is Newton's constant and $\ell_{\mbox{\tiny Pl}}$, is the Planck length. However, in
Section~\ref{sec:squeezed} we will work in units where  $ G = \ell_{\mbox{\tiny Pl}}^2 =1$ for consistency with previous references. In all case, we will restore powers 
 $\ell_{\mbox{\tiny Pl}}$ in final results. 
%
%
\section{Geodesic deviation fluctuations}
\label{sec:geodesic}
\begin{figure}[h]
	\centering
		\includegraphics[scale=0.20]{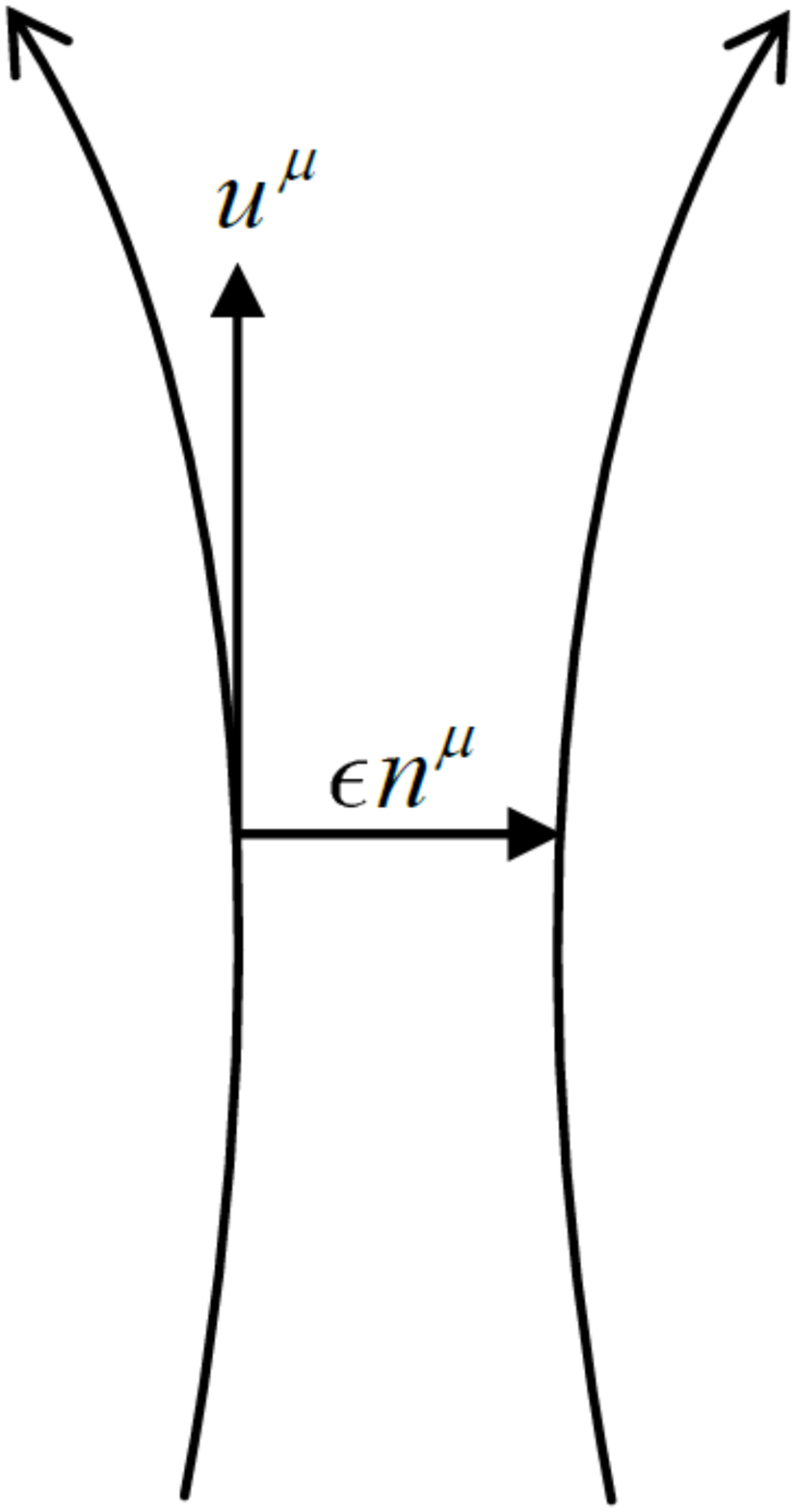}
	\caption{Timelike geodesics for two nearby falling particles, with four-velocity  $u^{\mu}$, and separation $\epsilon n^{\mu}$.   }
	\label{fig:Fig1_DEF}
\end{figure}

Let us consider two test particles whose worldlines are timelike geodesics infinitesimally separated, as represented in Fig.~\ref{fig:Fig1_DEF}, where $\epsilon \ll 1$.

The four-velocity, $u^{\mu}$, and the unit spacelike vector, $n^{\mu}$, are given by
\begin{equation}
u^{\mu} \equiv \frac{dx^{\mu}}{d\tau},
\label{eq:tangent_vector}
\end{equation}
\begin{equation}
n^{\mu} \equiv \frac{dx^{\mu}}{dn}\, , \qquad  n^{\mu}n_{\mu}=1.
\label{eq:deviation_vector}
\end{equation}
Note that we are considering the distance between the geodesics particles as a multiple of $n^{\mu}$.

The variation of the separation vector between two neighboring geodesics is described by the geodesic deviation equation,
\begin{equation}
\frac{D^{2}n^{\mu}}{d\tau^{2}}=-R_{\ \alpha\nu\lambda}^{\mu}u^{\alpha}n^{\nu}u^{\lambda},
\label{eq:geodesic_equation}
\end{equation}
where $\tau$ is the proper time, and $R_{\alpha\nu\lambda}^{\mu}$ is the Riemann tensor.
For particles on the neighboring geodesic, the relative acceleration per unit proper length is given by \cite{PhysRevD.80.124019}
\begin{equation}
\alpha \equiv n_{\mu}\frac{D^{2}n^{\mu}}{d\tau^{2}}=-R_{\mu\alpha\nu\lambda}n^{\mu}u^{\alpha}n^{\nu}u^{\lambda}.
\label{eq:relative_acceleration}
\end{equation}
Thus, if the particle starts at rest at proper time $\tau=0$, then at proper time $\tau=\tau_{0}$ we can define the relative velocity per unit proper length as
\begin{equation}
\upsilon \equiv n_{\mu}\frac{Dn^{\mu}}{d\tau}=-\int_{0}^{\tau_{0}}d\tau\ R_{\mu\alpha\nu\lambda}(\tau)\ n^{\mu}u^{\alpha}n^{\nu}u^{\lambda} \,,
\label{eq:relative_velocity}
\end{equation}
where the Riemann tensor depends upon the proper time along the geodesic.
Equation~(\ref{eq:relative_velocity}) defines the scalar velocity in the frame where $n^{\mu}=(0,\vec{n})$.
Note that $\upsilon$ is strictly a velocity per unit separation length, and hence has dimensions of inverse length in $c=1$ units.

Now, let us suppose that the spacetime geometry is subject to quantum fluctuations. In fact, given an ensemble of geodesics, measurements of the relative velocity 
along the same line will give different results. Therefore, we must take the expectation value of these measurements as well as the standard deviation. To do this, 
we will assume that the Riemann tensor is subject to quantum fluctuations which can be, in principle, active, passive or both.
We have to specify how the 4-vectors $u^{\mu}$ and $n^{\mu}$ behave under the fluctuations. The simplest assumption is that both $u^{\mu}$ and $n^{\mu}$ 
do not fluctuate to lowest order in the perturbations of spacetime. Physically, this is equivalent to assuming that both source and detector are located in a flat region, 
or both are rigidly attached to one another by non-gravitational forces. Finally, we assume that the perturbation is negligible at both source and detector.

The mean relative velocity of the particles, $\langle \upsilon \rangle$, is now obtained by averaging Eq.~(\ref{eq:relative_velocity}) as follows
\begin{equation}
\langle \upsilon \rangle =-\int_{0}^{\tau_{0}}d\tau\ \langle R_{\mu\alpha\nu\lambda}(\tau) \rangle\ n^{\mu}u^{\alpha}n^{\nu}u^{\lambda} \,.
\label{eq:mean_relative_velocity}
\end{equation}
The fluctuations around the mean trajectory in the direction of $n^{\mu}$ are described by
\begin{eqnarray}
\Delta\upsilon & = & \upsilon-\langle \upsilon \rangle\nonumber\\
 & = & -\int_{0}^{\tau_{0}}d\tau\ [R_{\mu\alpha\nu\lambda}(\tau)-\langle R_{\mu\alpha\nu\lambda}(\tau) \rangle]\ n^{\mu}u^{\alpha}n^{\nu}u^{\lambda}.
\label{eq:variation_relative_velocity}
\end{eqnarray}

Therefore, the variance of the relative velocity, $\langle (\Delta\upsilon)^{2} \rangle$, can be expressed as
\begin{eqnarray}
\langle (\Delta\upsilon)^{2} \rangle & = & \langle \upsilon^{2} \rangle-\langle \upsilon \rangle^{2}\nonumber\\
 & = & \int_{0}^{\tau_{0}}d\tau\int_{0}^{\tau_{0}}d\tau '\ C_{\alpha\lambda\mu\nu\gamma\delta\rho\sigma}(x,x')\ n^{\alpha}u^{\lambda}n^{\mu}u^{\nu}n^{\gamma}u^{\delta}n^{\rho}u^{\sigma},
\label{eq:standard_deviation_relative_velocity_correlation}
\end{eqnarray}
where the Riemann tensor correlation function, $C_{\alpha\lambda\mu\nu\gamma\delta\rho\sigma}(x,x')$, is given by
\begin{equation}
C_{\alpha\lambda\mu\nu\gamma\delta\rho\sigma}(x,x')=\langle R_{\alpha\lambda\mu\nu}(x) R_{\gamma\delta\rho\sigma}(x') \rangle-\langle R_{\alpha\lambda\mu\nu}(x) \rangle \langle R_{\gamma\delta\rho\sigma}(x') \rangle.
\label{eq:Riemann_tensor_correlation_function}
\end{equation}
This expression describes the fluctuations of the Riemann tensor. Here, the indices $\alpha\lambda\mu\nu$ refer to the spacetime point $x$ (which corresponds to the point $\tau$), 
while the indices $\gamma\delta\rho\sigma$ refer to the spacetime point $x'$ (which corresponds to the point $\tau'$).
Equation (\ref{eq:standard_deviation_relative_velocity_correlation}) is our key result for the geodesic deviation fluctuations and it applies to both active and passive fluctuations 
of the spacetime geometry. In what follows, we will evaluate the relative velocity dispersion given by Eq.~(\ref{eq:standard_deviation_relative_velocity_correlation}) for three
different sources of active fluctuations.
%
%
\section{Thermal graviton state}
\label{sec:thermal}
In this section, we will analyze the fluctuations produced by a thermal bath of gravitons, which may be created, for example, by the Hawking effect or 
cosmological particle production \cite{PhysRevD.70.044019,PhysRevD.94.084030}.
In this case, let us suppose that the spacetime geometry fluctuates in such a way that \cite{PhysRevD.74.024012}
\begin{equation}
\langle R_{\ \lambda\mu\nu}^{\alpha} \rangle=0,
\label{eq:average_value_Riemann_fluctuations}
\end{equation}
but
\begin{equation}
\langle R_{\ \lambda\mu\nu}^{\alpha} R_{\ \delta\rho\sigma}^{\gamma} \rangle \neq 0.
\label{eq:variance_Riemann_fluctuations}
\end{equation}
These two statements mean that we are neglecting the average spacetime curvature due to the bath of gravitons. Therefore, the average geometry corresponds to a
 flat Minkowski spacetime. Furthermore, since that we are dealing with a thermal quantum state at temperature $T$, the Riemann tensor correlation function can be written as
\begin{equation}
C_{\alpha\lambda\mu\nu\gamma\delta\rho\sigma}=\langle R_{\alpha\lambda\mu\nu}(x) R_{\gamma\delta\rho\sigma}(x') \rangle_{\beta},
\label{eq:Riemann_tensor_correlation_function_thermal}
\end{equation}
where $\langle R_{\alpha\lambda\mu\nu}(x) R_{\gamma\delta\rho\sigma}(x') \rangle_{\beta}$ is the thermal normal-ordered Riemann tensor two-point function, with $\beta=1/T$. Therefore, Eq.~(\ref{eq:standard_deviation_relative_velocity_correlation}) reduces to
\begin{equation}
\langle (\Delta\upsilon)^{2} \rangle=\int_{0}^{\tau_{0}}d\tau\int_{0}^{\tau_{0}}d\tau '\ \langle R_{\alpha\lambda\mu\nu}(x) R_{\gamma\delta\rho\sigma}(x') \rangle_{\beta}\ n^{\alpha}u^{\lambda}n^{\mu}u^{\nu}n^{\gamma}u^{\delta}n^{\rho}u^{\sigma}.
\label{eq:standard_deviation_relative_velocity_thermal}
\end{equation}

Let us choose the case where both source and detector are initially at rest with respect to one another and to the bath of gravitons. This choice is such that
\begin{equation}
u^{\mu}=(1,0,0,0),
\label{eq:four_velocity_TT_gauge}
\end{equation}
\begin{equation}
n^{\mu}=(0,1,0,0),
\label{eq:separation_vector_TT_gauge}
\end{equation}
where we have assumed that the particles are separated in the $x$-direction. 
Thus, substituting Eqs.~(\ref{eq:four_velocity_TT_gauge})-(\ref{eq:separation_vector_TT_gauge}) into Eq.~(\ref{eq:standard_deviation_relative_velocity_thermal}), we obtain
\begin{eqnarray}
\langle (\Delta\upsilon)^{2} \rangle & = & \int_{0}^{\tau_{0}}d\tau\int_{0}^{\tau_{0}}d\tau '\ \langle R_{xtxt}(x) R_{xtxt}(x') \rangle_{\beta}\ n^{x}u^{t}n^{x}u^{t}n^{x}u^{t}n^{x}u^{t}\nonumber\\
 & = & \int_{0}^{\tau_{0}}d\tau\int_{0}^{\tau_{0}}d\tau '\ \langle R_{txtx}(x) R_{txtx}(x') \rangle_{\beta},
\label{eq:standard_deviation_relative_velocity_thermal_TT}
\end{eqnarray}
where we have used the symmetry and cyclic properties of the Riemann tensor, namely,
\begin{equation}
R_{\alpha\lambda\mu\nu}=-R_{\lambda\alpha\mu\nu}=-R_{\alpha\lambda\nu\mu}.
\label{eq:Riemann_tensor_properties}
\end{equation}

Now, we introduce the thermal Riemann tensor two-point function which was constructed from the vacuum two-point function via the Matsubara method 
(see \cite{PhysRevD.80.124019} and references therein). It is given by
\begin{equation}
\langle R_{txtx}(x) R_{txtx}(x') \rangle_{\beta}=\frac{1}{4}(\partial_{t}^{4}-2\partial_{t}^{2}\partial_{x}^{2}+\partial_{x}^{4})D_{\beta},
\label{eq:Riemann_tensor_two-point}
\end{equation}
with
\begin{equation}
D_{\beta}=\frac{1}{4\pi^{2}}\sum_{n=-\infty}^{+\infty\ '}\frac{1}{(\Delta\vec{x})^{2}-(\Delta t+in\beta)^{2}},
\label{eq:D_beta}
\end{equation}
where
\begin{equation}
\partial_{t}\partial_{t'}=-\partial_{t}^{2},
\label{eq:partial_t}
\end{equation}
\begin{equation}
\Delta\vec{x}=\vec{x}-\vec{x}',
\label{eq:Delta_x}
\end{equation}
\begin{equation}
\Delta t=t-t'.
\label{eq:Delta_t}
\end{equation}
In the last summation, the prime denotes that we  have removed the $n=0$  term, which is the vacuum contribution.

Now we will examine the relative velocity dispersion in one space dimension, that is, we may choose $\Delta y=\Delta z=0$. 
Then, we can write Eq.~(\ref{eq:D_beta}) as
\begin{equation}
D_{\beta}=\frac{1}{4\pi^{2}}\sum_{n=-\infty}^{+\infty\ '}\frac{1}{(\Delta x)^{2}-(\Delta t+in\beta)^{2}}\,.
\label{eq:D_beta_x}
\end{equation}
Next assume that the two particles both start at rest in our frame of reference , so we may use $d\tau = dt$ in Eq.~(\ref{eq:standard_deviation_relative_velocity_thermal_TT}),
which becomes
\begin{eqnarray}
\langle (\Delta\upsilon)^{2} \rangle=I_{t}+I_{tx}+I_{x},
\label{eq:standard_deviation_relative_velocity_thermal_TT_I}
\end{eqnarray}
with
\begin{equation}
I_{t}=\int_{0}^{t_{0}}dt\int_{0}^{t_{0}}dt'\ \biggl(\frac{1}{4}\partial_{t}^{4}D_{\beta}\biggr),
\label{eq:I_t}
\end{equation}
\begin{equation}
I_{tx}=\int_{0}^{t_{0}}dt\int_{0}^{t_{0}}dt'\ \biggl(-\frac{1}{2}\partial_{t}^{2}\partial_{x}^{2}D_{\beta}\biggr),
\label{eq:I_tx}
\end{equation}
\begin{equation}
I_{x}=\int_{0}^{t_{0}}dt\int_{0}^{t_{0}}dt'\ \biggl(\frac{1}{4}\partial_{x}^{4}D_{\beta}\biggr),
\label{eq:I_x}
\end{equation}
where $t_{0}$ is the flight time, that is, the interaction time between the particles and the thermal bath.

We are interested in the real part of $D_{\beta}$. Thus, we may make the replacement
\begin{equation}
\sum_{n=-\infty}^{+\infty\ '}=2\sum_{n=1}^{+\infty},
\label{eq:sum_replacement}
\end{equation}
and take
\begin{equation}
\Re(D_{\beta})=\frac{1}{2\pi^{2}}\sum_{n=1}^{+\infty}G,
\label{eq:D_beta_real}
\end{equation}
with
\begin{equation}
G=\Re\biggl[\frac{1}{(\Delta x)^{2}-(\Delta t+in\beta)^{2}}\biggr],
\label{eq:G_TBG}
\end{equation}
where $\Re$ denotes the real part. After that, we can evaluate Eq.~(\ref{eq:standard_deviation_relative_velocity_thermal_TT_I}) by using an algebraic manipulation program. 
However, the final expression is so long that no insight is gained by writing it out.

Next, we will assume that $\Delta x$ is small compared to $\Delta t$ and/or $\beta$, and hence can be ignored. We may compute the relative velocity between the two test 
particles taking the following limits in the derivatives:
\begin{eqnarray}
\frac{1}{4}\left.\partial_{t}^{4}D_{\beta}\right|_{x \rightarrow x'} & = & \frac{1}{2\pi^{2}}\sum_{n=1}^{+\infty}\biggl(\frac{1}{4}\left.\partial_{t}^{4}G\right|_{x \rightarrow x'}\biggr)\nonumber\\
& = & \frac{1}{2\pi^{2}}\sum_{n=1}^{+\infty}\Re\biggl[-\frac{30}{(\Delta t+in\beta)^{6}}\biggr],
\label{eq:D_t}
\end{eqnarray}
\begin{eqnarray}
-\frac{1}{2}\left.\partial_{t}^{2}\partial_{x}^{2}D_{\beta}\right|_{x \rightarrow x'} & = & \frac{1}{2\pi^{2}}\sum_{n=1}^{+\infty}\biggl(-\frac{1}{2}\left.\partial_{t}^{2}\partial_{x}^{2}G\right|_{x \rightarrow x'}\biggr)\nonumber\\
& = & \frac{1}{2\pi^{2}}\sum_{n=1}^{+\infty}\Re\biggl[\frac{20}{(\Delta t+in\beta)^{6}}\biggr],
\label{eq:D_tx}
\end{eqnarray}
\begin{eqnarray}
\frac{1}{4}\left.\partial_{x}^{4}D_{\beta}\right|_{x \rightarrow x'} & = & \frac{1}{2\pi^{2}}\sum_{n=1}^{+\infty}\biggl(\frac{1}{4}\left.\partial_{x}^{4}G\right|_{x \rightarrow x'}\biggr)\nonumber\\
& = & \frac{1}{2\pi^{2}}\sum_{n=1}^{+\infty}\Re\biggl[-\frac{6}{(\Delta t+in\beta)^{6}}\biggr].
\label{eq:D_x}
\end{eqnarray}
Thus, we can write the thermal normal-ordered Riemann tensor two-point function as
\begin{equation}
\langle R_{txtx}(x) R_{txtx}(x') \rangle_{\beta}=\frac{1}{2\pi^{2}}\sum_{n=1}^{+\infty}\Re\biggl[-\frac{16}{(\Delta t+in\beta)^{6}}\biggr].
\label{eq:Riemann_tensor_two-point_thermal_case}
\end{equation}
Therefore, substituting Eqs.~(\ref{eq:I_t})-(\ref{eq:D_x}) into Eq.~(\ref{eq:standard_deviation_relative_velocity_thermal_TT_I}), we obtain
\begin{eqnarray}
\langle (\Delta\upsilon)^{2} \rangle & = & 
\frac{1}{2\pi^{2}}\sum_{n=1}^{+\infty}\biggl[\frac{8}{5n^{4}\beta^{4}}-\frac{8}{5(t_{0}^{2}+n^{2}\beta^{2})^{2}}+
\frac{64t_{0}^{2}}{5(t_{0}^{2}+n^{2}\beta^{2})^{3}}-\frac{64t_{0}^{4}}{5(t_{0}^{2}+n^{2}\beta^{2})^{4}}\biggr].
\label{eq:standard_deviation_relative_velocity_thermal_TT_I_result}
\end{eqnarray}
At this point, we can analyze the limits in which the time of observation $t_{0}$ is large compared to the thermal parameter $\beta$, and vice versa.

\subsection{Case 1: \texorpdfstring{$t_{0} \ll \beta$}{t0 << beta} (short flight time or low temperature)}
If $t_{0} \ll \beta$, we have
\begin{equation}
\langle (\Delta\upsilon)^{2} \rangle \sim 
\frac{1}{2\pi^{2}}\sum_{n=1}^{+\infty}\frac{16 t_{0}^{2}}{n^{6}\beta^{6}}=\frac{8\pi^{4}t_{0}^{2}}{945\beta^{6}}=
\frac{256 \pi^{5} \ell_{\mbox{\tiny Pl}}^{2} \,t_{0}^{2}}{945\beta^{6}}.
\label{eq:standard_deviation_relative_velocity_thermal_TT_I_result_t_final_small}
\end{equation}
Here the rms relative velocity  is given by
\begin{equation}
(\Delta\upsilon)_{\mbox{\scriptsize rms}}=\frac{16 \pi^{5/2} \ell_{\mbox{\tiny Pl}}\, t_0}{3 \sqrt{105}\beta^{3}},
\label{eq:standard_deviation_relative_velocity_thermal_TT_I_result_rms_small}
\end{equation}
which grows linearly with the flight time. Recall that, following the convention in Ref.~\cite{PhysRevD.80.124019}, we set $32 \pi \ell_{\mbox{\tiny Pl}}^2 =1$
in this section.

\subsection{Case 2: \texorpdfstring{$t_{0} \gg \beta$}{t0 >> beta} (long flight time or high temperature)}
In the limit when $t_{0} \gg \beta$, that is, in the observationally reasonable limit where the wavelength of the gravitational waves is negligible with respect to the flight time of the particles after the interaction with the thermal bath of gravitons, the expression reduces to
\begin{equation}
\langle (\Delta\upsilon)^{2} \rangle \sim \frac{1}{2\pi^{2}}\sum_{n=1}^{+\infty}\frac{8}{5n^{4}\beta^{4}}=\frac{2\pi^{2}}{225\beta^{4}}=\frac{64 \pi^{3}\ell_{\mbox{\tiny Pl}}^{2}}{225\beta^{4}},
\label{eq:standard_deviation_relative_velocity_thermal_TT_I_result_final}
\end{equation}
where $\ell_{\mbox{\tiny Pl}}$ is the Planck length. In this case, the rms relative velocity dispersion approaches a constant, namely,
\begin{equation}
(\Delta\upsilon)_{\mbox{\scriptsize rms}}=\frac{8 \pi^{3/2} \ell_{\mbox{\tiny Pl}}}{15\beta^{2}}.
\label{eq:standard_deviation_relative_velocity_thermal_TT_I_result_rms}
\end{equation}

\subsection{Position fluctuations}
\begin{figure}
	\centering
		\includegraphics[scale=0.20]{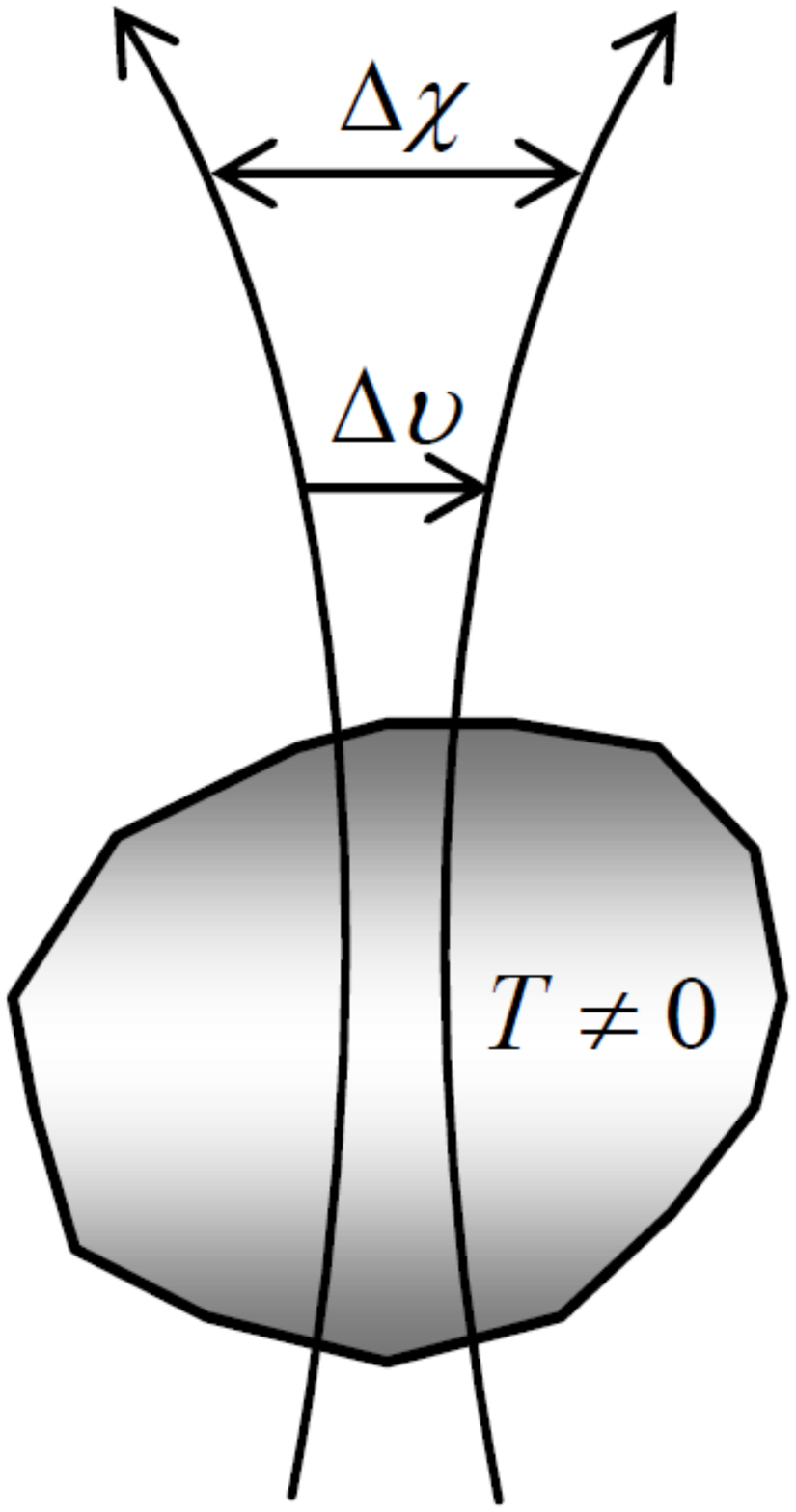}
	\caption{Distance and velocity between two nearby timelike geodesics after interaction with the thermal bath of gravitons.}
	\label{fig:Fig2_DEF}
\end{figure}

From the previous calculations, note that the relative velocity dispersion is not zero, that is, the nearby timelike geodesics are affected by the gravitons. 
In this way, we are interested in computing the relative distance dispersion between the two test particles after their interaction with the thermal bath of gravitons.
The mean squared distance fluctuation in the $x$-direction, as represented in Fig.~\ref{fig:Fig2_DEF}, can be calculated as follows:
\begin{equation}
\langle (\Delta \chi)^{2} \rangle = \int_{0}^{\mathcal{T}}dt_{1}\int_{0}^{t_{1}}dt\int_{0}^{\mathcal{T}}dt'_{1}\int_{0}^{t'_{1}}dt'\ \langle R_{txtx}(x) R_{txtx}(x') \rangle_{\beta}.
\label{eq:standard_deviation_position_thermal_TT}
\end{equation}
Note that $\chi$ is a fractional distance, and hence is dimensionless in arbitrary units.
 
Following the same procedure used to obtain Eq.~(\ref{eq:standard_deviation_relative_velocity_thermal_TT_I_result}), namely, setting 
$\Delta x,\Delta y,\Delta z \approx 0$ in the denominator of the Riemann tensor correlation function, which means $\Delta \vec{x} \apprle \Delta t$, the relative distance 
dispersion is given by
\begin{equation}
\langle (\Delta \chi)^{2} \rangle = \frac{1}{2\pi^{2}}\sum_{n=1}^{+\infty}\biggl[\frac{4}{15n^{2}\beta^{2}}+\frac{4t^{2}}{5n^{4}\beta^{4}}+
\frac{4(3t^{4}-6t^{2}n^{2}\beta^{2}-n^{4}\beta^{4})}{15(t^{2}+n^{2}\beta^{2})^{3}}\biggr],
\label{eq:standard_deviation_position_thermal_TT_result}
\end{equation}
where we have set $t=\mathcal{T} > t_{0}$, which means that $t$ is the measurement time, i.e, the total time given by the sum of the flight time and the time
elapsed after the interaction.

In the $t \ll \beta$ limit, the relative distance dispersion and its root-mean-square value are given, respectively, by
\begin{equation}
\langle (\Delta \chi)^{2} \rangle \sim \frac{1}{2\pi^{2}}\sum_{n=1}^{+\infty}\frac{4t^{4}}{n^{6}\beta^{6}}=\frac{64 \pi^{5} \ell_{\mbox{\tiny Pl}}^{2} \, t^4}{945\beta^{6}},
\label{eq:standard_deviation_position_thermal_TT_case2}
\end{equation}
\begin{equation}
(\Delta \chi)_{\mbox{\scriptsize rms}}=\frac{8 \pi^{5/2} \ell_{\mbox{\tiny Pl}}\,  t^{2}  }{3 \sqrt{105}\beta^{3}}.
\label{eq:standard_deviation_position_thermal_TT_case2_rms}
\end{equation}
On the other hand, for the $t \gg \beta$ limit, we have
\begin{equation}
\langle (\Delta \chi)^{2} \rangle \sim \frac{1}{2\pi^{2}}\sum_{n=1}^{+\infty}\frac{4t^{2}}{5n^{4}\beta^{4}}=\frac{32 \pi^{3} \ell_{\mbox{\tiny Pl}}^2 \, t^2}{225\beta^{4}},
\label{eq:standard_deviation_position_thermal_TT_case1}
\end{equation}
\begin{equation}
(\Delta \chi)_{\mbox{\scriptsize rms}}=\frac{4 \sqrt{2} \pi^{3/2}  \ell_{\mbox{\tiny Pl}}  \, t} {15\beta^{2}}.
\label{eq:standard_deviation_position_thermal_TT_case1_rms}
\end{equation}
From Eqs.~(\ref{eq:standard_deviation_relative_velocity_thermal_TT_I_result_rms}), (\ref{eq:standard_deviation_relative_velocity_thermal_TT_I_result_rms_small}), 
(\ref{eq:standard_deviation_position_thermal_TT_case1_rms}), and (\ref{eq:standard_deviation_position_thermal_TT_case2_rms}), we have
\begin{equation}
(\Delta \chi)_{\mbox{\scriptsize rms}}=\frac{(\Delta \upsilon)_{\mbox{\scriptsize rms}}}{2t_{0}}t^{2} \qquad (\mbox{for}\ t \ll \beta).
\label{eq:standard_deviation_position_thermal_TT_case2_rms_compared}
\end{equation}
\begin{equation}
(\Delta \chi)_{\mbox{\scriptsize rms}}=\frac{(\Delta \upsilon)_{\mbox{\scriptsize rms}}}{\sqrt{2}}t \qquad (\mbox{for}\ t \gg \beta),
\label{eq:standard_deviation_position_thermal_TT_case1_rms_compared}
\end{equation}
Therefore, we find in both cases a form of gravitational wave memory effect (see Ref.~\cite{PhysRevD.96.064013} and references therein), such that
\begin{equation}
(\Delta \chi)_{\mbox{\scriptsize rms}} \sim (\Delta\upsilon)_{\mbox{\scriptsize rms}}\ t.
\label{eq:standard_deviation_position_thermal_TT_case1_rms_comparison}
\end{equation}
This means that the timelike geodesics are modified due to presence of the thermal bath of gravitons, as expected. This modification is reflected in the distance 
between the pair of particles and  can in principle become large as $t$ grows. despite the smallness of the Planck length.

\section{Gravitons in a squeezed quantum state}
\label{sec:squeezed}

In this section we will consider a spacetime region filled with gravitons in a squeezed state which produce quantum fluctuations on test particle geodesics. In principle, this squeezed state
could be due to quantum creation of  gravitons in a background gravitational field, as for example, in the course of a cosmological expansion or in the Hawking process of 
black hole evaporation \cite{ClassQuantumGrav.6.L161,PhysRevD.42.3413,PhysRevD.46.645}. 
The squeezed state is represented by $|\alpha,\zeta\rangle$, where $\alpha$ and $\zeta$ are the displacement and squeezed parameters, respectively. 
Consider a gravitational plane wave mode in a squeezed state. The 
normal-ordered Riemann tensor correlation function can be expressed as \cite{PhysRevD.74.024012}
\begin{equation}
: C_{\alpha\lambda\mu\nu\gamma\delta\rho\sigma}(x,x') :\ = 4(\ell_{[\alpha}A_{\lambda][\mu}\ell_{\nu]})(\ell_{[\gamma}A_{\delta][\rho}\ell_{\sigma]})F(x,x'),
\label{eq:Riemann_tensor_correlation_function_GSS}
\end{equation}
with
\begin{eqnarray}
F(x,x') & = & [\cosh(2r)-1]\cos[\ell_{\epsilon}(x^{\epsilon}-x'^{\epsilon})]-\sinh(2r)\cos[\ell_{\epsilon}(x^{\epsilon}+x'^{\epsilon})+\theta],
\label{eq:F1_GSS}
\end{eqnarray}
where $\ell^{\epsilon}=(\omega_{g},\ell^{x},\ell^{y},\ell^{z})$ is the specific wave vector of the excited mode, $\omega_{g}$ is the frequency, $A_{\mu\nu}$ is the polarization tensor, 
and the parameters $r,\theta$ are defined such that $\zeta=r \mbox{e}^{i\theta}$.
We have used the following convention for the antisymmetrized tensor,
\begin{equation}
T_{[\mu\nu]}=\frac{1}{2}(T_{\mu\nu}-T_{\nu\mu})\,.
\label{eq:skew-simmetric}
\end{equation}
In this section only, we follow the convention of Ref.~ \cite{PhysRevD.74.024012}, and use units in which $ \ell_{\mbox{\tiny Pl}} =1$.
Substituting Eqs.~(\ref{eq:Riemann_tensor_correlation_function_GSS}) and (\ref{eq:F1_GSS}) into Eq.~(\ref{eq:standard_deviation_relative_velocity_correlation}), 
the relative velocity dispersion for test particles subjected to gravitons in a squeezed state is given by
\begin{equation}
\langle (\Delta\upsilon)^{2} \rangle = 
4(\ell_{[\alpha}A_{\lambda][\mu}\ell_{\nu]})(\ell_{[\gamma}A_{\delta][\rho}\ell_{\sigma]})n^{\alpha}u^{\lambda}n^{\mu}u^{\nu}n^{\gamma}u^{\delta}n^{\rho}u^{\sigma}f_{1}(\omega_{g},t_{0}),
\label{eq:standard_deviation_relative_velocity_GSS}
\end{equation}
where the function $f_{1}(\omega_{g},t_{0})$ is calculated from Eq.~(\ref{eq:standard_deviation_relative_velocity_thermal_TT}), again with  $d\tau = dt$,  leading to
\begin{eqnarray}
f_{1}(\omega_{g},t_{0}) & = & \int_{0}^{\tau_{0}}d\tau\int_{0}^{\tau_{0}}d\tau'\ F(x,x')\nonumber\\
 & = & \int_{0}^{t_{0}}dt\int_{0}^{t_{0}}dt'\ \biggl.\{[\cosh(2r)-1]\cos[\omega_{g}(t-t')+\ell_{x}(x-x')]\nonumber\\
 &   & -\sinh(2r)\cos[\omega_{g}(t+t')+\ell_{x}(x+x')+\ell_{y}(y+y')+\ell_{z}(z+z')+\theta]\biggr.\}\nonumber\\
 & = & \frac{4}{\omega_{g}^2}\sin^{2}\biggl(\frac{\omega_{g}t_{0}}{2}\biggr)\{2\sinh^{2}(r)\cos[\ell_{x}(x-x')]\nonumber\\
 &   & -\sinh(2r)\cos[\omega_{g}t_{0}+\ell_{x}(x+x')+\ell_{y}(y+y')+\ell_{z}(z+z')+\theta]\},
\label{eq:f1_GSS}
\end{eqnarray}
where $t_{0}$ is the flight time, the interaction time between the geodesic particles and the quantum state under consideration.

In this last result, we have chosen $\Delta y=\Delta z=0$, that is, we are again assuming that $\Delta y$ and $\Delta z$ are small. It is worth calling attention to the fact that 
$f_{1}(\omega_{g},t_{0})$, and hence $\langle (\Delta\upsilon)^{2} \rangle$, are independent of the displacement parameter $\alpha$. 
Therefore, the fluctuations depend only on the squeezing parameter $\zeta$ in such a way that $\zeta=0$ (a coherent state) induces no fluctuations:
\begin{equation*}
r=0 \Rightarrow \zeta=0 \Rightarrow f_{1}=0 \Rightarrow \langle (\Delta\upsilon)^{2} \rangle=0 \qquad \mbox{(classical wave)}.
\label{eq:no_fluctuations}
\end{equation*}

We will assume that the gravitational wave mode is in the transverse tracefree (TT) gauge, in which the gravitational perturbations have only spatial components $h_{ij}$, 
satisfying  $\partial^{i}h_{ij}=0$ and $h_{i}^{i}=0$.
In fact, this choice is only a matter of convenience, since our results were obtained from the Riemann tensor which is gauge invariant. 
Thus, the first vector product of Eq.~(\ref{eq:standard_deviation_relative_velocity_GSS}) is given by
\begin{eqnarray}
(\ell_{[\alpha}A_{\lambda][\mu}\ell_{\nu]})n^{\alpha}u^{\lambda}n^{\mu}u^{\nu} & = & (\ell_{[x}A_{t][x}\ell_{t]})n^{x}u^{t}n^{x}u^{t}\nonumber\\
 & = & \frac{1}{4}(\ell_{x}A_{tx}\ell_{t}-\ell_{x}A_{tt}\ell_{x}-\ell_{t}A_{xx}\ell_{t}+\ell_{t}A_{xt}\ell_{x})\nonumber\\
 & = & -\frac{1}{4}(\ell_{t})^{2}A_{xx}\nonumber\\
 & = & -\frac{1}{4}\omega_{g}^{2}A_{+},
\label{eq:standard_deviation_relative_velocity_GSS_TT_first}
\end{eqnarray}
where the polarization tensor $A_{\mu\nu}$ is given by
\begin{eqnarray}
A_{\mu\nu}=\left(
\begin{array}{cccc}
	0 & 0      & 0       & 0\\
	0 & A_{xx} & A_{xy}  & 0\\
	0 & A_{xy} & -A_{xx} & 0\\
	0 & 0      & 0       & 0
\end{array}
\right)
=
\left(
\begin{array}{cccc}
	0 & 0      & 0       & 0\\
	0 & A_{+} & A_{\times}  & 0\\
	0 & A_{\times} & -A_{+} & 0\\
	0 & 0      & 0       & 0
\end{array}
\right),
\label{eq:polarization_1_GSS}
\end{eqnarray}
which is obviously traceless and purely spatial:
\begin{equation}
A_{0\nu}=0,\quad \eta^{\mu\nu}A_{\mu\nu}=0,\quad A_{3\nu}=0.
\label{eq:polarization_2_GSS}
\end{equation}

Therefore, there exists a nonzero effect on the relative velocity dispersion due to gravitons in a squeezed state, which depends on the $(+)$ polarization as well as upon
position. It is given by
\begin{eqnarray}
\langle (\Delta\upsilon)^{2} \rangle & = & 4\biggl(-\frac{1}{4}\omega_{g}^{2}A_{+}\biggr)\biggl(-\frac{1}{4}\omega_{g}^{2}A_{+}\biggr)f_{1}(\omega_{g},t_{0})\nonumber\\
& = & \frac{1}{4}\omega_{g}^{4}A_{+}^{2}f_{1}(\omega_{g},t_{0}).
\label{eq:standard_deviation_relative_velocity_GSS_TT}
\end{eqnarray}
In the $t_{0} \rightarrow 0$ limit, we can expand Eq.~(\ref{eq:standard_deviation_relative_velocity_GSS_TT}) for fixed $r$ in order to find
\begin{eqnarray}
\langle (\Delta\upsilon)^{2} \rangle & \sim & \frac{1}{4}\omega_{g}^{4}A_{+}^{2} \{2\sinh^{2}(r)\cos[\ell_{x}(x-x')]\nonumber\\
& & -\sinh(2r)\cos[\ell_{x}(x+x')+\ell_{y}(y+y')+\ell_{z}(z+z')+\theta]\}t_{0}^{2}.
\label{eq:standard_deviation_relative_velocity_GSS_TT_limit}
\end{eqnarray}
%
%
\subsection{Classical time-dependence}
In this subsection ,we examine the expectation value of the relative velocity, $\langle \upsilon \rangle$, which is given in terms of the first order contribution in the 
Riemann tensor fluctuations, $\langle R_{\alpha\lambda\mu\nu}(x) \rangle$. This quantity gives the classical time dependent variation since it depends only upon the 
displacement parameter, $\alpha$. If $\zeta=0$ the squeezed state becomes a coherent state, which can describe a classical wave.

In order to evaluate $\langle \upsilon \rangle$, we can do a single integration of the expectation value of the Riemann tensor over the proper time $d\tau$ by using Eq.~(\ref{eq:mean_relative_velocity}). However, we want to calculate $\langle \upsilon \rangle^{2}$ directly from Eq.~(\ref{eq:standard_deviation_relative_velocity_correlation}). 
Then, the squared mean change in relative velocity can be written as
\begin{equation}
\langle \upsilon \rangle^{2} = \int_{0}^{\tau_{0}}d\tau\int_{0}^{\tau_{0}}d\tau '\ \langle R_{\alpha\lambda\mu\nu}(x) \rangle \langle R_{\gamma\delta\rho\sigma}(x') \rangle n^{\alpha}u^{\lambda}n^{\mu}u^{\nu}n^{\gamma}u^{\delta}n^{\rho}u^{\sigma}.
\label{eq:standard_change_relative_velocity_correlation_CTD}
\end{equation}
The right-hand side of the Riemann tensor correlation function for gravitons in a squeezed state is given by \cite{PhysRevD.74.024012}
\begin{equation}
\langle R_{\alpha\lambda\mu\nu}(x) \rangle \langle R_{\gamma\delta\rho\sigma}(x') \rangle = 4 \sum_{\ell} (\ell_{[\alpha}A_{\lambda][\mu}\ell_{\nu]})(\ell_{[\gamma}A_{\delta][\rho}\ell_{\sigma]})G(x,x'),
\label{eq:Riemann_tensor_correlation_function_GSS_CTD}
\end{equation}
where
\begin{equation}
G(x,x')=\alpha^{2}\mbox{e}^{i\ell_{\epsilon}(x^{\epsilon}+x'^{\epsilon})}+(\alpha^{*})^{2}\mbox{e}^{-i\ell_{\epsilon}(x^{\epsilon}+x'^{\epsilon})}+2|\alpha|^{2}\cos[\ell_{\epsilon}(x^{\epsilon}-x'^{\epsilon})].
\label{eq:F2_GSS_CTD}
\end{equation}
Thus, following the same procedure used to obtain Eq.~(\ref{eq:f1_GSS}), and performing the integration of $G(x,x')$, we get
\begin{eqnarray}
g_{1}(\omega_{g},t_{0}) & = & \int_{0}^{\tau_{0}}d\tau\int_{0}^{\tau_{0}}d\tau'\ G(x,x')\nonumber\\
 & = & \int_{0}^{t_{0}}dt\int_{0}^{t_{0}}dt'\ \biggl.\{2|\alpha|^{2}\cos[\omega_{g}(t-t')+\ell_{x}(x-x')]\nonumber\\
 &   & +(\alpha^{*})^{2}\mbox{e}^{-i[\omega_{g}(t+t')+\ell_{x}(x+x')+\ell_{y}(y+y')+\ell_{z}(z+z')]}+\alpha^{2}\mbox{e}^{i[\omega_{g}(t+t')+\ell_{x}(x+x')+\ell_{y}(y+y')+\ell_{z}(z+z')]}\biggr.\}\nonumber\\
 & = & -\frac{\mbox{e}^{-i[2\omega_{g}t_{0}+\ell_{x}(x+x')+\ell_{y}(y+y')+\ell_{z}(z+z')]}(\mbox{e}^{i\omega_{g}t_{0}}-1)^{2}}{\omega_{g}^2}\nonumber\\
 &   & \times\{(\alpha^{*})^{2}+\alpha^{2}\mbox{e}^{2i[\omega_{g}t_{0}+\ell_{x}(x+x')+\ell_{y}(y+y')+\ell_{z}(z+z')]}\nonumber\\
 &   & +2|\alpha|^{2}\mbox{e}^{i[\omega_{g}t_{0}+\ell_{x}(x+x')+\ell_{y}(y+y')+\ell_{z}(z+z')]}\cos[\ell_{x}(x-x')]\}.
\label{eq:f3_GSS_CTD}
\end{eqnarray}

Therefore, the classical time dependent variation of the relative velocity for a single mode is characterized by
\begin{eqnarray}
\langle \upsilon \rangle^{2} & = & 4(\ell_{[\alpha}A_{\lambda][\mu}\ell_{\nu]})(\ell_{[\gamma}A_{\delta][\rho}\ell_{\sigma]}) n^{\alpha}u^{\lambda}n^{\mu}u^{\nu}n^{\gamma}u^{\delta}n^{\rho}u^{\sigma}g_{1}(\omega_{g},t_{0})\nonumber\\
 & = & 4\biggl(-\frac{1}{4}\omega_{g}^{2}A_{+}\biggr)\biggl(-\frac{1}{4}\omega_{g}^{2}A_{+}\biggr)g_{1}(\omega_{g},t_{0})\nonumber\\
 & = & \frac{1}{4}\omega_{g}^{4}A_{+}^{2}g_{1}(\omega_{g},t_{0}).
\label{eq:final_change_relative_velocity_correlation_CTD}
\end{eqnarray}
%
%
\subsection{Special case: Transverse gravitational waves}
Here, we will analyze the special case of transversely propagating gravity waves. These waves propagate with wave vector given by $\ell^{\mu}=\omega_{g}(1,0,0,1)$, while the 
test particles continue to have 4-vectors given by $u^{\mu}=(1,0,0,0)$ and $n^{\mu}=(0,1,0,0)$.
Then, for a gravitational wave propagating in the $z$-direction, we just need to set $\ell_{x}=\ell_{y}=0$ into Eq.~(\ref{eq:f1_GSS}) in order to define the function $f_{2}(\omega_{g},t_{0})$ as
\begin{equation}
f_{2}(\omega_{g},t_{0})=\frac{4}{\omega_{g}^2}\sin^{2}\biggl(\frac{\omega_{g}t_{0}}{2}\biggr)\{2\sinh^{2}(r)-\sinh(2r)\cos[\omega_{g}t_{0}+\omega_{g}(z+z')+\theta]\}.
\label{eq:f2_GSS_z}
\end{equation}
Therefore, in the special case of transverse gravitational waves, the relative velocity dispersion is given by
\begin{equation}
\langle (\Delta\upsilon)^{2} \rangle = \frac{1}{4}\omega_{g}^{4}A_{+}^{2}f_{2}(\omega_{g},t_{0}).
\label{eq:standard_deviation_relative_velocity_GSS_z}
\end{equation}
From Eqs.~(\ref{eq:standard_deviation_relative_velocity_GSS_TT}) and (\ref{eq:standard_deviation_relative_velocity_GSS_z}), we conclude that the fluctuations in the relative velocity depend on the degree of squeezing, measured by the parameter $\zeta$.

On the other hand, the classical time dependent variaton of the relative velocity is given by
\begin{equation}
\langle \upsilon \rangle^{2}=\frac{1}{4}\omega_{g}^{4}A_{+}^{2}g_{2}(\omega_{g},t_{0}),
\label{eq:final_change_relative_velocity_correlation_CTD_z}
\end{equation}
where the function $g_{2}(\omega_{g},t_{0})$ is defined in a similar way as $g_{1}(\omega_{g},t_{0})$, when $\ell_{x}=\ell_{y}=0$, namely,
\begin{eqnarray}
g_{2}(\omega_{g},t_{0}) & = & -\frac{\mbox{e}^{-i\omega_{g}[2t_{0}+(z+z')]}(\mbox{e}^{i\omega_{g}t_{0}}-1)^{2}}{\omega_{g}^{2}}\nonumber\\
& & \times\{(\alpha^{*})^{2}+2|\alpha|^{2}\mbox{e}^{i\omega_{g}[t_{0}+(z+z')]}+\alpha^{2}\mbox{e}^{2i\omega_{g}[t_{0}+(z+z')]}\}.
\label{eq:f4_GSS_CTD}
\end{eqnarray}
Note that both functions $g_{1}(\omega_{g},t_{0})$ and $g_{2}(\omega_{g},t_{0})$ depend on the displacement parameter, $\alpha$, but are independent of the squeeze parameter, 
$r$. Therefore, we can say the same for Eqs.~(\ref{eq:final_change_relative_velocity_correlation_CTD}) and (\ref{eq:final_change_relative_velocity_correlation_CTD_z}). 
Furthermore, in the $\alpha=0$ limit, we have that $g_{1}(\omega_{g},t_{0})=g_{2}(\omega_{g},t_{0})=0$, which means a coherent state ($r=0$ and $\alpha \neq 0$) exhibits 
regular time variation but does not fluctuate. In fact, from Eq.~(\ref{eq:f1_GSS}) we can see that $\langle (\Delta\upsilon)^{2} \rangle=0$ for $r=0$.
%
%
\subsection{Estimating \texorpdfstring{$\langle (\Delta\upsilon)^{2} \rangle$}{<(Delta v)>} from the value of the stress tensor}
In this subsection, we will estimate the order of magnitude of $\langle (\Delta\upsilon)^{2} \rangle$ in the squeezed vacuum state when $\alpha=0$ and $r \gg 1$. In order to do this, 
we will assume the $(+)$ polarization for the gravitational waves, which implies $A_{\times}=0$. These assumptions lead to $\langle \upsilon \rangle^{2}=0$ and, therefore, we have 
$\langle (\Delta\upsilon)^{2} \rangle=\langle \upsilon^{2} \rangle$.
Thus, from Eqs.~(\ref{eq:f1_GSS}) and (\ref{eq:standard_deviation_relative_velocity_GSS_TT}), with a suitable choice of $\theta$, in a situation where many modes are excited, we find the following asymptotic behavior for large $r$:
\begin{eqnarray}
\langle (\Delta\upsilon)^{2} \rangle & \approx & \frac{1}{4}\omega_{g}^{4}\frac{8\pi}{\omega_{g}V}\frac{V}{(2\pi)^{3}}(\Delta\ell_{x})(\Delta\ell_{y})(\Delta\ell_{z})\frac{2\mbox{e}^{2r}}{\omega_{g}^{2}}\nonumber\\
& = & \frac{\omega_{g}\mbox{e}^{2r}}{2\pi^{2}}(\Delta\ell_{x})(\Delta\ell_{y})(\Delta\ell_{z}),
\label{eq:standard_deviation_relative_velocity_GSS_estimate}
\end{eqnarray}
where the contribution from the $(+)$ polarization is given by $A_{+}=\sqrt{8\pi/\omega_{g}V}$. In the latter result, we have summed the modes when the density of states is large, namely,
\begin{equation}
\sum_{\ell} \rightarrow \frac{V}{(2\pi)^{3}} \int d^{3}\ell=\frac{V}{(2\pi)^{3}}(\Delta\ell_{x})(\Delta\ell_{y})(\Delta\ell_{z}),
\label{eq:density_states}
\end{equation}
where $V$ is the quantization volume.
Now, from the effective stress tensor in the linearized theory, the vacuum energy density for large $r$ is given by \cite{PhysRevD.74.024012}
\begin{equation}
:T_{00}:\ \approx \frac{\omega_{g}\mbox{e}^{2r}}{32\pi^{3}}(\Delta\ell_{x})(\Delta\ell_{y})(\Delta\ell_{z}).
\label{eq:vacuum_energy_density}
\end{equation}
Then, substituting Eq.~(\ref{eq:vacuum_energy_density}) into Eq.~(\ref{eq:standard_deviation_relative_velocity_GSS_estimate}), the relative velocity dispersion can be expressed as
\begin{equation}
\langle (\Delta\upsilon)^{2} \rangle \approx \frac{\omega_{g}\mbox{e}^{2r}}{2\pi^{2}}\frac{32\pi^{3}}{\omega_{g}\mbox{e}^{2r}}:T_{00}:\ =16\pi:T_{00}:\ell_{\mbox{\tiny Pl}}^{2}.
\label{eq:standard_deviation_relative_velocity_GSS_estimate_final}
\end{equation}
In this case, the rms relative velocity dispersion is given by
\begin{equation}
(\Delta\upsilon)_{\mbox{\scriptsize rms}} \approx \ell_{\mbox{\tiny Pl}}\,\sqrt{:T_{00}:}.
\label{eq:standard_deviation_relative_velocity_GSS_estimate_rms}
\end{equation}

In order to estimate this value, let us suppose a closure energy density given by $:T_{00}:\ =10^{8}$ cm$^{-4}$. Therefore, taking into account the Planck length $\ell_{\mbox{\tiny Pl}}=10^{-33}$ cm, we obtain an estimate of the fractional relative speed
\begin{equation}
(\Delta\upsilon)_{\mbox{\scriptsize rms}} \approx 10^{-29} {\rm cm}^{-1}.
\label{eq:standard_deviation_relative_velocity_GSS_estimate_rms_final}
\end{equation}
This implies that two test particles separated by a distance of one light year, or $10^{18} {\rm cm}$, would acquire an actual relative speed of about $10^{-19} c =0.3 \,{\rm cm/s}$, 
which is very small. However, in the early universe where graviton density could be much larger, the effect could increase.

%
%
\subsection{Position fluctuation}
In order to compute the relative distance dispersion for two geodesic particles subject to the gravitons in a squeezed state, we follow the same procedure used in the thermal case, 
and perform two more integrals of the function $F(x,x')$:
\begin{equation}
\langle (\Delta\chi)^{2} \rangle = 4(\ell_{[\alpha}A_{\lambda][\mu}\ell_{\nu]})(\ell_{[\gamma}A_{\delta][\rho}\ell_{\sigma]})n^{\alpha}u^{\lambda}n^{\mu}u^{\nu}n^{\gamma}u^{\delta}n^{\rho}u^{\sigma}f_{3}(\omega_{g},t_{0}),
\label{eq:standard_deviation_relative_distance_GSS}
\end{equation}
where the function $f_{3}(\omega_{g},t_{0})$ is given by
\begin{eqnarray}
f_{3}(\omega_{g},\mathcal{T}) & = & \int_{0}^{\mathcal{T}}d\tau_{1}\int_{0}^{\tau_{1}}d\tau\int_{0}^{\mathcal{T}}d\tau'_{1}\int_{0}^{\tau'_{1}}d\tau'\ F(x,x')\nonumber\\
& = & \int_{0}^{\mathcal{T}}dt_{1}\int_{0}^{t_{1}}dt\int_{0}^{\mathcal{T}}dt'_{1}\int_{0}^{t'_{1}}dt'\ \biggl.\{[\cosh(2r)-1]\cos[\omega_{g}(t-t')+\ell_{x}(x-x')]\nonumber\\
&   & -\sinh(2r)\cos[\omega_{g}(t+t')+\ell_{x}(x+x')+\ell_{y}(y+y')+\ell_{z}(z+z')+\theta]\biggr.\},
\label{eq:f3_GSS}
\end{eqnarray}
such that
\begin{eqnarray}
f_{3}(\omega_{g},t) & = & \frac{1}{\omega_{g}^{4}}\{2 \sinh ^2(r) \cos [l_{x} (x-x')] [t^2 \omega_{g} ^2-2 t \omega_{g}  \sin (t \omega_{g} )-2 \cos (t \omega_{g} )+2]\nonumber\\
&   & +\sinh (2 r) \{t^2 \omega_{g} ^2 \cos [\theta +l_{y}(y+y')+l_{z}(z+z')+l_{x} (x+x')]\nonumber\\
&   & +2 t \omega_{g}  \sin [\theta +l_{y}(y+y')+l_{z}(z+z')+l_{x} (x+x')]\nonumber\\
&   & -2 t \omega_{g}  \sin [\theta +l_{y}(y+y')+l_{z}(z+z')+l_{x} (x+x')+t \omega_{g} ]\nonumber\\
&   & +2 \cos [\theta +l_{y}(y+y')+l_{z}(z+z')+l_{x} (x+x')+t \omega_{g} ]\nonumber\\
&   & -\cos [\theta +l_{y}(y+y')+l_{z}(z+z')+l_{x} (x+x')+2 t \omega_{g} ]\nonumber\\
&   & -\cos [\theta +l_{y}(y+y')+l_{z}(z+z')+l_{x} (x+x')]\}\},
\label{eq:f3_GSS_t}
\end{eqnarray}
where we have set $t=\mathcal{T} > t_{0}$, which means that $t$ is the total time.

Therefore, the relative distance dispersion is given by
\begin{equation}
\langle (\Delta\chi)^{2} \rangle = \frac{1}{4}\omega_{g}^{4}A_{+}^{2}f_{3}(\omega_{g},t).
\label{eq:standard_deviation_relative_distance_GSS_full}
\end{equation}
In the $t \rightarrow 0$ limit, we can expand Eq.~(\ref{eq:standard_deviation_relative_distance_GSS_full}) for fixed $r$ in order to get
\begin{eqnarray}
\langle (\Delta\chi)^{2} \rangle & \sim & \frac{1}{4}\omega_{g}^{4}A_{+}^{2} \biggl\{\frac{1}{2}\sinh^{2}(r)\cos[\ell_{x}(x-x')]\nonumber\\
& & -\frac{1}{4}\sinh(2r)\cos[\ell_{x}(x+x')+\ell_{y}(y+y')+\ell_{z}(z+z')+\theta]\biggr\}t^{4}.
\label{eq:standard_deviation_relative_distance_GSS_TT_limit}
\end{eqnarray}
From Eqs.~(\ref{eq:standard_deviation_relative_velocity_GSS_TT_limit}) and (\ref{eq:standard_deviation_relative_distance_GSS_TT_limit}), we have
\begin{eqnarray}
\langle (\Delta\chi)^{2} \rangle = \frac{\langle (\Delta\upsilon)^{2} \rangle}{4}t^{2}.
\label{eq:standard_deviation_relative_distance_GSS_TT_limit_compared}
\end{eqnarray}
The root-mean-square value is given by
\begin{equation}
(\Delta \chi)_{\mbox{\scriptsize rms}}=\frac{(\Delta \upsilon)_{\mbox{\scriptsize rms}}}{2}t.
\label{eq:standard_deviation_position_GSS_TT_limit_rms_compared}
\end{equation}
Therefore, we conclude that
\begin{equation}
(\Delta \chi)_{\mbox{\scriptsize rms}} \sim (\Delta\upsilon)_{\mbox{\scriptsize rms}}\ t.
\label{eq:standard_deviation_position_GSS_TT_limit_rms_comparison}
\end{equation}
This is the same behavior as for the thermal case, given by Eq.~(\ref{eq:standard_deviation_position_thermal_TT_case1_rms_comparison}).
%
%
\section{Graviton vacuum state}
\label{sec:vacuum}

In this section, we will deal with fluctuations of the Riemann tensor in the graviton vacuum state in linearized quantum gravity. This state must approximate 
a corresponding state in full quantum gravity in a suitable limit, and exhibits nontrivial fluctuation effects. 
From Eq.~(\ref{eq:standard_deviation_relative_velocity_correlation}), the relative velocity dispersion can be expressed as
\begin{equation}
\langle (\Delta\upsilon)^{2} \rangle = \int_{0}^{\tau_{0}} d\tau \int_{0}^{\tau_{0}} d\tau'\ C_{txtxtxtx}(x,x').
\label{eq:standard_deviation_relative_velocity_vacuum}
\end{equation}
where the Riemann tensor correlation function, $C_{txtxtxtx}(x,x')$, is given in terms of the vacuum two-point function, namely,
\begin{equation}
C_{txtxtxtx}(x,x')=\langle R_{txtx}(x) R_{txtx}(x') \rangle=\frac{1}{4}(\partial_{t}^{4}-2\partial_{t}^{2}\partial_{x}^{2}+\partial_{x}^{4})D,
\label{eq:Riemann_tensor_two-point_vacuum}
\end{equation}
with
\begin{equation}
D=\frac{1}{4\pi^{2}[(\Delta \vec{x})^{2}-(\Delta t)^{2}]}.
\label{eq:D_vacuum}
\end{equation}
Substituting Eqs.~(\ref{eq:partial_t})-(\ref{eq:Delta_t}) into Eq.~(\ref{eq:Riemann_tensor_two-point_vacuum}), we can write the full expression of the Riemann tensor 
correlation function in the graviton vacuum state as
\begin{eqnarray}
C_{txtxtxtx}(\Delta t,\Delta x,\Delta y,\Delta z) & = & 
\frac{4}{\pi^{2}} \biggl[\frac{(\Delta x^{2}-\Delta t^{2})^{2}+(\Delta y^{2}+\Delta z^{2})^{2}}{(\Delta x^{2}+\Delta y^{2}+\Delta z^{2}-\Delta t^{2})^{5}}\nonumber\\
& & -\frac{4(\Delta x^{2}-\Delta t^{2})(\Delta y^{2}+\Delta z^{2})}{(\Delta x^{2}+\Delta y^{2}+\Delta z^{2}-\Delta t^{2})^{5}}\biggr],
\label{eq:Riemann_tensor_two-point_vacuum_state}
\end{eqnarray}
where $\Delta t=t-t'$, $\Delta x=x-x'$, $\Delta y=y-y'$, and $\Delta z=z-z'$. 

Here, we have an undesirable singularity when we try to compute the expectation value of $(\Delta\upsilon)^{2}$. Then, we need to adopt an operational approach which 
 expresses spacetime averages of the correlation function as finite integrals. We choose the approach developed in Ref.~ \cite{PhysRevD.70.064032}  to compute the 
 fluctuations in the focusing of a bundle of geodesics.
Therefore, in order to get a finite value for the integral given by Eq.~(\ref{eq:standard_deviation_relative_velocity_vacuum}), we will sample over a spacetime volume 
(the interior of a world tube) defined by a bundle of geodesics.  The history of a wave packet is formed by integrating along the time and averaging in space. To do this, we replace the 
integrations on the proper time in Eq.~(\ref{eq:standard_deviation_relative_velocity_vacuum}) by four-dimensional spacetime integrations \cite{PhysRevD.72.105010}
\begin{equation}
\langle (\Delta\upsilon)^{2} \rangle = \int_{-\infty}^{+\infty} d^{4}x\ f(x) \int_{-\infty}^{+\infty} d^{4}x'\ f(x')\ C_{txtxtxtx}(x,x'),
\label{eq:standard_deviation_relative_velocity_vacuum_bundle}
\end{equation}
where $f(x)$ is the sampling function. Here we will first average in space, and then integrate in time.

For the spatial averaging, we use a Lorentzian sampling function of width $\phi$  in each of the rectangular coordinates $x,y,z$ and $x',y',z'$,
\begin{equation}
g_{\mbox{\tiny L}}(u,\phi)=\frac{\phi}{\pi(u^{2}+\phi^{2})},
\label{eq:Gaussian_function}
\end{equation}
so that
\begin{equation}
\int_{-\infty}^{+\infty} du\ g_{\mbox{\tiny L}}(u,\phi)=1.
\label{eq:normalization_Gaussian_function}
\end{equation}
This has the effect of averaging over a spatial scale of order $\phi$. Therefore, the Riemann tensor correlation function, averaged over the spatial directions, may be defined  by
\begin{eqnarray}
\hat{C}(t-t',b) & = & \int_{-\infty}^{+\infty} d^{3}x\ f(x) \int_{-\infty}^{+\infty} d^{3}x'\ f(x')\ C_{txtxtxtx}(x,x')\nonumber\\
& = & \int_{-\infty}^{+\infty} dx\ g_{\mbox{\tiny L}}(x,\phi) \int_{-\infty}^{+\infty} dy\ g_{\mbox{\tiny L}}(y,\phi) \int_{-\infty}^{+\infty} dz\ g_{\mbox{\tiny L}}(z,\phi)\nonumber\\
&   & \times \int_{-\infty}^{+\infty} dx'\ g_{\mbox{\tiny L}}(x',\phi) \int_{-\infty}^{+\infty} dy'\ g_{\mbox{\tiny L}}(y',\phi) \int_{-\infty}^{+\infty} dz'\ g_{\mbox{\tiny L}}(z',\phi)\nonumber\\
&   & \biggl. \times\ C_{txtxtxtx}(\Delta t,\Delta x,\Delta y,\Delta z) \biggr.\nonumber\\
& = & \frac{4[3b^{4}+6b^{2}(t-t')^{2}-(t-t')^{4}]}{\pi^{2}[3b^{2}+(t-t')^{2}]^{5}}\,.
\label{eq:time_correlation_function_deviation_relative_position_vacuum_bundle}
\end{eqnarray}
Here we have used the following identity
\begin{equation}
\int_{-\infty}^{+\infty} dx\ g_{\mbox{\tiny L}}(x,\phi) \int_{-\infty}^{+\infty} dx'\ g_{\mbox{\tiny L}}(x',\phi)\ F(x-x') = \int_{-\infty}^{+\infty} d\Delta x\ g_{\mbox{\tiny L}}(\Delta x,b)\ F(\Delta x),
\label{eq:identity_Lorentzian_function_2}
\end{equation}
with $b=2\phi$. Note that the lightcone singularity present in Eq.~(\ref{eq:Riemann_tensor_two-point_vacuum_state}), is no longer present in $\hat{C}(t-t',b)$. We may interpret the
latter quantity as an acceleration correlation function which has been averaged in space, but not in time.

\subsection{Direct time integration}

Because $\hat{C}(t-t',b)$ is finite for all values of its arguments, so long as $b\not=0$, one option seems to be to integrate it directly in time to find the associated velocity and
position fluctuations. Define a velocity correlation function obtained by direct time integration by
\begin{equation}
\langle \upsilon(t_{1})\upsilon(t_{2}) \rangle_{\mbox{\tiny DTI}} = \int_{0}^{t_{1}} dt \int_{0}^{t_{2}} dt'  \hat{C}(t-t')\,,
\label{eq:DTI-v-corr}
\end{equation}
and the associated velocity variance at time $t_0$ by
\begin{equation}
\langle (\Delta\upsilon(t))^{2} \rangle_{\mbox{\tiny DTI}}  = \langle \upsilon(t_0)\upsilon(t_0) \rangle_{\mbox{\tiny DTI}} \,.
\end{equation}
The latter quantity is found to be
\begin{equation}
\langle (\Delta\upsilon)^2 \rangle_{\mbox{\tiny DTI}}  = \frac{16 \,t_0  \,\ell_{\mbox{\tiny Pl}}^2
[27b^{5}t_{0}+60b^{3}t_{0}^{3}+9bt_{0}^{5}+7\sqrt{3}(3b^{2}+t_{0}^{2})^{3}\arctan(t_{0}/\sqrt{3}b)]}{81 \pi \, b^{5}(3b^{2}+t_{0}^{2})^{3}} \,.
\label{eq:v2DTI}
\end{equation}
In the limit that $t_0$ becomes large for fixed $b$, we find
\begin{equation}
\langle (\Delta\upsilon)^{2} \rangle_{\mbox{\tiny DTI}}  \sim \frac{56 \,\sqrt{3} \,\ell_{\mbox{\tiny Pl}}^{2}}{ 81\,  b^{5}}  t_0   -
\frac{64 \ell_{\mbox{\tiny Pl}}^{2}}{27 \pi b^{4}}+\mathcal{O}\biggl(\frac{1}{t_0}\biggl)^{2}\,.
\label{eq:DTI-v2-lim1}
\end{equation}
One may also find the associated position fluctuations by further time integrations:
\begin{equation}
\langle (\Delta\chi)^{2} \rangle_{\mbox{\tiny DTI}}  = \int_{0}^{t_0}d t_{1}  \int_{0}^{t_0}d t_{2}   \langle \upsilon(t_{1})\upsilon(t_{2}) \rangle_{\mbox{\tiny DTI}} \,,
\label{eq:standard_deviation_position_vacuum}
\end{equation}
and find 
\begin{equation}
\langle (\Delta\chi)^{2} \rangle_{\mbox{\tiny DTI}}  \sim \frac{56 \ell_{\mbox{\tiny Pl}}^{2}\, t_0^{3} }{81 \sqrt{3}  b^{5}}
\label{eq:position_fluctuation_vacuum_t_infty}
\end{equation}
in the limit of large $t_0$.

These results are  puzzling, because they implies that the mean squared velocity of the particle grows linearly in time. This is only possible if there is an external energy source.
It is useful to examine a somewhat different limit. Let $b = c\, t_0$, where $c > 0$ is a constant. Now Eq.~(\ref{eq:v2DTI}) take the form
 \begin{equation}
\langle (\Delta\upsilon)^2 \rangle_{\mbox{\tiny DTI}} =  \frac{K}{t_0^4}\,,
\label{eq:DTIb}
\end{equation}
where $K$ is a constant. Now $ \langle (\Delta\upsilon)^2 \rangle_{\mbox{\tiny DTI}}  \rightarrow 0$ as $t_0 \rightarrow \infty$. Thus if both $b$ and $t_0$ become large together, then the
 velocity variance vanishes. At this point, it is unclear whether the linear growth found in Eq.~(\ref{eq:v2DTI}) is due to holding $b$ fixed, or to the sudden time switching
 used in the direct time integration approach.

.

\subsection{Lorentzian Time Integration}

We next adopt an indirect way of integrating in time, which we call Lorentzian time integration. It involves a dimensionless 
Lorentzian function given by
\begin{equation}
\bar{g}_{\mbox{\tiny L}}(u,\varphi)=\frac{\varphi^{2}}{\pi(u^{2}+\varphi^{2})}\,.
\label{eq:Lorentzian_function}
\end{equation}
This function has the following property
\begin{equation}
\int_{-\infty}^{+\infty} du\ \bar{g}_{\mbox{\tiny L}}(u,\varphi)=\varphi,
\label{eq:normalization_Lorentzian_function}
\end{equation}
so $\varphi$ is the effective interval of integration. That is, 
$\int_{-\infty}^{+\infty} du\ \bar{g}_{\mbox{\tiny L}}(u,\varphi)\, F(u)$ 
is an integral of $F(u)$ over an interval of order  $\varphi$ centered about $u =0$. The advantages of this approach are that the integral can be finite even if
$F(u)$ has a singularity somewhere in the range of integration, and it avoids sudden switching.

In this subsection, we will use Lorentzian time integration to study velocity fluctuations.
Thus, we integrate $\hat{C}(t-t',b)$ using two functions of the form of Eq.~(\ref{eq:Lorentzian_function}), and define the velocity variance as
\begin{eqnarray}
\langle   (\Delta\upsilon(t))^{2}  \rangle_{\mbox{\tiny LTI}} & = & 
\int_{-\infty}^{+\infty} dt\ \bar{g}_{\mbox{\tiny L}}(t,\varphi) \int_{-\infty}^{+\infty} dt'\ \bar{g}_{\mbox{\tiny L}}(t',\varphi)\ \hat{C}(t-t',b)\nonumber\\
& = & \frac{a}{4}\int_{-\infty}^{+\infty} d\tau\ \bar{g}_{\mbox{\tiny L}}(\tau,a)\ \hat{C}(\tau,b) \,,
\label{eq:LTI-v2}
\end{eqnarray}
where $a=2\varphi$. In the last step, we used the fact that
\begin{equation}
\int_{-\infty}^{+\infty} dt\ \bar{g}_{\mbox{\tiny L}}(t,\varphi) \int_{-\infty}^{+\infty} dt'\ \bar{g}_{\mbox{\tiny L}}(t',\varphi)\ F(t-t')=
\frac{a}{4}\int_{-\infty}^{+\infty} d\tau\ \bar{g}_{\mbox{\tiny L}}(\tau,a)\ F(\tau) \,.
\label{eq:identity_Lorentzian_function}
\end{equation} 

The integrand in the second line of Eq.~(\ref{eq:LTI-v2}) has first order poles at $\tau= \pm ia$ and fifth order poles 
at $\tau=\pm i \sqrt{3}\,b$. The integral may be performed by contour integration, with the result
\begin{eqnarray}
\langle   (\Delta\upsilon(t))^{2}  \rangle_{\mbox{\tiny LTI}}  &=&
\frac{2 \ell_{\mbox{\tiny Pl}}^2 \, a^2}{81 \pi \, b^5  \,(a^2-3 b^2)^5} \, (7 \sqrt{3} a^9-108 \sqrt{3} a^7 b^2 + 594 \sqrt{3} a^5 b^4 \\
&+& 1296 a^4 b^5  - 4860 \sqrt{3} a^3 b^6+7776 a^2 b^7+1215 \sqrt{3} a  b^8-3888 b^9 ) \,.
\label{eq:LTI-v2b}
\end{eqnarray} 
In the limit that $a$ becomes large for fixed $b$, we have
\begin{equation}
\langle   (\Delta\upsilon(t))^{2}  \rangle_{\mbox{\tiny LTI}} \sim \frac{14 \sqrt{3} \, a \,\ell_{\mbox{\tiny Pl}}^{2}}{81 \, b^{5}} \,.
\label{eq:LTI-v2c}
\end{equation} 
Given that the duration of the time integration is proportional to $a$, this is essentially the same result as in Eq.~(\ref{eq:DTI-v2-lim1}), with the velocity variance
growing linearly in the flight time. In fact, if we set $a = 4 \pi t_0$, the two asymptotic forms are identical.
We can also consider the limit where $a$ and $b$ are proportional to one another: set $b =c \, a$, so Eq.~(\ref{eq:LTI-v2b}) takes the form
\begin{equation}
\langle (\Delta\upsilon)^2 \rangle_{\mbox{\tiny LTI}} =  \frac{K'}{a^4}\,,
\label{eq:LTIb}
\end{equation}
for some constant $K'$. 
Now $ \langle (\Delta\upsilon)^2 \rangle_{\mbox{\tiny LTI}}  \rightarrow 0$ as $a \rightarrow \infty$.

Both Eqs.~(\ref{eq:LTI-v2c}) and (\ref{eq:LTIb}) are in qualitative agreement with the corresponding results,   Eqs.~(\ref{eq:DTI-v2-lim1}) and (\ref{eq:DTIb}), found using direct
time integration. This indicates that the linear growth of $\langle   (\Delta\upsilon(t))^{2}  \rangle$ in time is not an artifact of sudden temporal switching. However, the velocity 
variance does not grow when both the flight time and spatial scale increase together. This result is suggests that we should examine more general space and time averagings.

\subsection{Averaging over World Tubes of increasing Width}

In both of the previous subsections, the spatial scale $b$ was a constant, which means that we were averaging over the history of a bundle of rays with a fixed spatial 
cross section. However, more realistic beams tend to spread in width as they propagate. Now we explore an averaging method which describes this spreading. Return 
to the Riemann tensor correlation function, Eq.~(\ref{eq:Riemann_tensor_two-point_vacuum_state}). Now we average it with Lorentzians  of width $b$ in $(x,y,z)$,
but width $b'$ in $(x',y',z')$. However, this is equivalent to averaging with Lorentzians of width $b+b'$ in each of $\Delta x$, $\Delta y$, and $\Delta z$, because
of the identity
\begin{equation}
\int_{-\infty}^{+\infty} dx\ g_{\mbox{\tiny L}}(x,b) \int_{-\infty}^{+\infty} dx'\ g_{\mbox{\tiny L}}(x',b')\ F(x-x') = 
\int_{-\infty}^{+\infty} d\Delta x\ g_{\mbox{\tiny L}}(\Delta x,b+b')\ F(\Delta x) \,.
\label{eq:identity_Lorentzian_function_3}
\end{equation}
Thus, we may define
\begin{eqnarray}
\hat{C}(t-t',b,b')  &=&  \int_{-\infty}^{+\infty} dx \,dy\,dz \,g_{\mbox{\tiny L}}(x,b) \, g_{\mbox{\tiny L}}(y,b) \, g_{\mbox{\tiny L}}(z,b)\nonumber\\
& \times& \int_{-\infty}^{+\infty} dx' \, dy' \, dz'  \,g_{\mbox{\tiny L}}(x',b') \, g_{\mbox{\tiny L}}(y',b')\,  g_{\mbox{\tiny L}}(z',b') \;
 C_{txtxtxtx}(\Delta t,\Delta x,\Delta y,\Delta z)  \nonumber\\
 &=&  \int_{-\infty}^{+\infty} d\Delta x \,d\Delta y\,d\Delta z \,g_{\mbox{\tiny L}}(\Delta x,b+b') \, g_{\mbox{\tiny L}}(\Delta y,b+b') \, g_{\mbox{\tiny L}}(\Delta z,b+b') \nonumber\\
 &\times& C_{txtxtxtx}(\Delta t,\Delta x,\Delta y,\Delta z)  \,.
\label{eq:b-bp-ave}
\end{eqnarray}
This may be evaluated with the result
\begin{equation}
\hat{C}(t-t',b,b') = \frac{4[3(b+b')^{4}+6(b+b')^{2}(t-t')^{2}-(t-t')^{4}]}{\pi^{2}[3(b+b')^{2}+(t-t')^{2}]^{5}}\,.
\label{eq:b-bp-ave2}
\end{equation}

So far, $b$ and $b'$ have been constants, but they may be functions of time without changing any of the above analysis. Let them be linear functions given by
\begin{equation}
b = b(t) = c\, t + b_0 \quad {\rm and } \quad b' = b'(t') = c\, t' + b_0\,, 
\end{equation}
where $c > 0$ and $b_0 > 0$ are constants. These functions describe a bundle of rays which starts with a nonzero width, which then grows linearly in time as 
the rays propagate. Now the variable width averaged acceleration correlation function becomes
\begin{equation}
\hat{C}_{vw}(t,t') = \hat{C}(t-t',b(t),b'(t')) = 
\frac{\{4[3[c(t+t') +2b_0]^{4}+6[c(t+t') +2b_0]^{2}(t-t')^{2}-(t-t')^{4}\}}{\pi^{2}\{3[c(t+t') +2b_0]^{2}+(t-t')^{2}\}^{5}}\,.
\label{eq:b-bp-ave3}
\end{equation}

\begin{figure}[ht]
	\centering
		\includegraphics[scale=0.40]{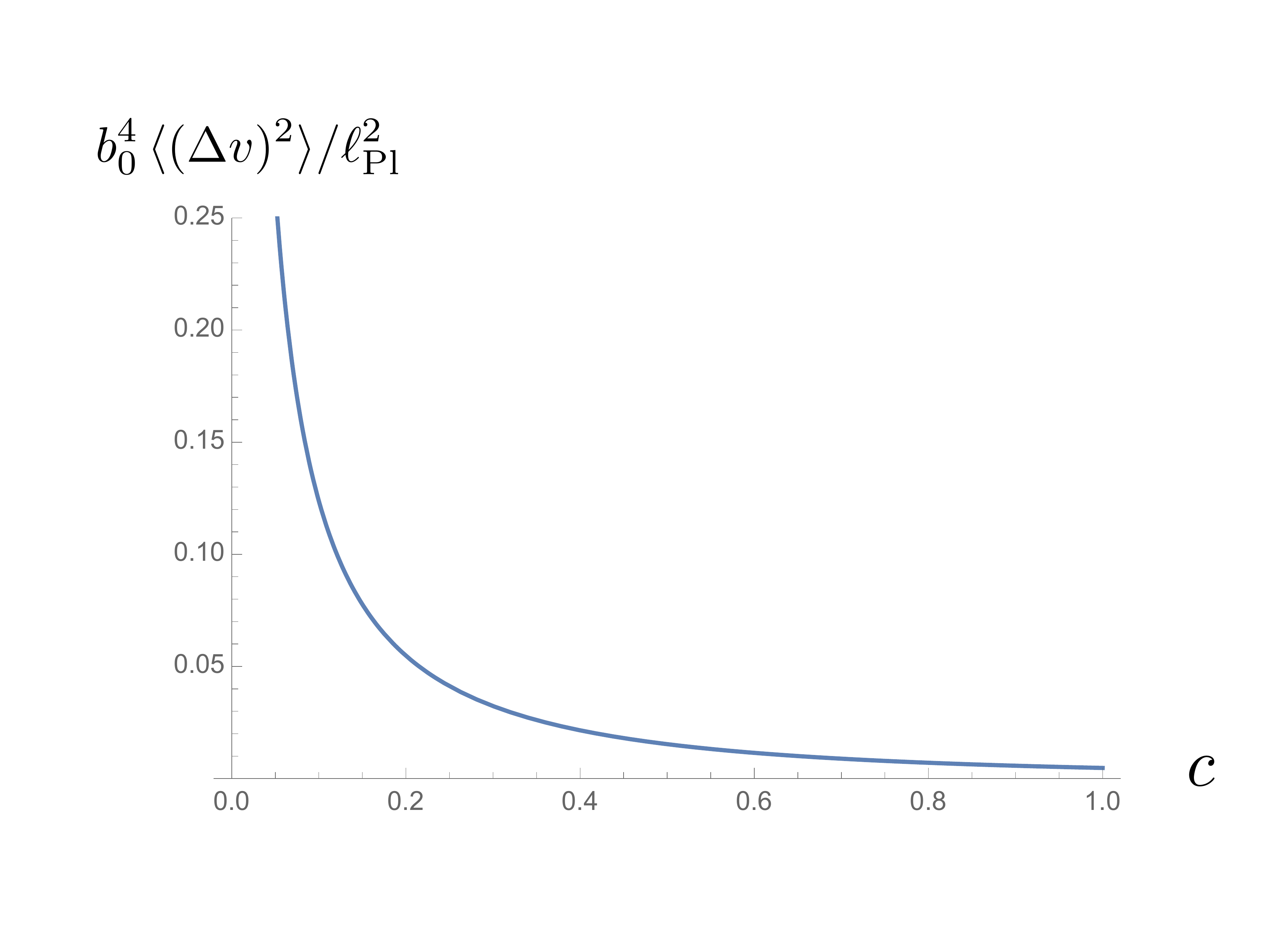}
	\caption{The velocity variance for the case of variable width spatial sampling is plotted as a function of the parameter $c$.  }
	\label{fig:v2-plot}
\end{figure}

The velocity variance of obtained from an integral upon $t$ and $t'$ of $\hat{C}_{vw}(t,t')$. In the limit of a long flight time, this variance becomes
\begin{equation}
\langle (\Delta\upsilon)^2 \rangle = \int_0^\infty dt \int_0^\infty dt' \, \hat{C}_{vw}(t,t') \,.
\label{eq:vw-v2}
\end{equation}
This integral is finite so long as both $c$ and $b_0$ are nonzero. This is most easily seen by transforming to polar coordinates, defined by $t = \tau\, \sin\theta$ and
$t' = \tau \, \cos\theta$, so
\begin{equation}
\langle (\Delta\upsilon)^2 \rangle = \int_0^\infty d\tau \int_0^{\pi/2} d\theta \, \tau \,\hat{C}_{vw}(\tau,\theta) \,.
\label{eq:vw-v2b}
\end{equation}
The integrand is finite as $\tau \rightarrow 0$ so long as $b_0 > 0$. As $\tau \rightarrow \infty$, the integrand falls as $1/\tau^5$ for all $\theta$ if $c >0$, and hence the integral
converges at the upper limit. We can also now understand why we found $\langle (\Delta\upsilon)^2 \rangle$ growing with increasing flight time in the two previous subsections. 
Both of those cases correspond to $c=0$ in the present notation. If $c=0$, the integrand in Eq.~(\ref{eq:vw-v2b}) grows for large $\tau$ if $\theta =\pi/4$, which is the
$t = t'$ line. Note that on dimensional 
grounds, $\langle (\Delta\upsilon)^2 \rangle \propto {b_0}^{-4}$. The integral in Eq.~(\ref{eq:vw-v2}) may be evaluated numerically as a function of the parameter $c$, and the
result is plotted in Fig.~\ref{fig:v2-plot}.

In summary, we have found that averaging over a geodesic bundle with a fixed spatial cross section leads to a mean squared velocity which grows linearly in time. This
requires an external energy source to supply the added kinetic energy to the particles. However, if the cross section grows linearly in time, as would be the case for a
diverging beam of particles, then the  mean squared velocity approaches a constant value. Furthermore, this asymptotic value is very small unless the initial cross section
is close to the Planck scale. In other contexts, the lack of secular growth of vacuum fluctuation effects can be linked to anti-correlations~\cite{PhysRevD.72.105010}. It 
is of interest to explore whether similar anti-correlations exist here as well. This is a topic for future research. 

\section{Conclusions}
\label{sec:final}

In this work, we have analyzed the effects of fluctuations of the spacetime geometry on the motion of test particles using the geodesic deviation equation. Just as a classical
gravitational field leads to tidal acceleration and changes in the relative velocities of test particles, a fluctuating gravitational field leads to fluctuations in these relative
velocities and consequently fluctuations in the relative separations of the particles. We treat the geodesic deviation equation as a Langevin equation, which may be integrated
to express the relative velocity and position variances as integrals of a Riemann tensor correlation function Here we have considered fluctuations around an average flat 
spacetime background produced by linear quantum gravity effects. Thus we are dealing with active fluctuations of the dynamical degrees of freedom of gravity, as opposed 
to passive fluctuations driven by a matter stress tensor.  The source of the spacetime geometry fluctuations could be either a bath of gravitons, or the graviton vacuum
fluctuations. We have consider both a thermal bath of gravitons, and a bath of gravitons in a squeezed vacuum state. As expected, the velocity and position variances tend
to be very small, and are suppressed by the square of the ratio of the Planck length to a characteristic length scale of the system.  In the case of a thermal graviton bath,
the variance of the relative velocity approaches a constant at late time, but root-mean squared position fluctuation grows linearly in time. This can be interpreted as a version
of the gravitational memory effect~\cite{PhysRevD.96.064013}. 

The discussion of graviton vacuum fluctuation effects, given in Section~\ref{sec:vacuum}, requires averaging over both space and time to produce finite results. We can view
this as averaging over a world tube which describes the history of a set of test particles. In the case where the spatial width of the bundle of geodesics is held constant, we find
that the mean squared relative velocity grows linearly with the flight time. This seems to require an external energy source to maintain the constant width.  However, it also
raises the interesting possibility of enhanced quantum gravity effects for long flight times. However, we also find that if the spatial width is allowed to grow, even if very slowly,
as the particles propagate, then the mean squared relative velocity approaches a constant.

There seem to be some subtle effects of spacetime geometry fluctuations in linearized quantum gravity which may elucidate the effects to be expected in a complete quantum
gravity theory.

%
%
\begin{acknowledgments}
This work was supported in part by the U.S. National Science Foundation under Grant PHY-1607118. H.S.V. is funded by the Brazilian research agencies CNPq (research Project No. 140612/2014-9) and CAPES (PDSE Process No. 88881.133092/2016-01). V.B.B. is partially supported by the CNPq through the research Project No. 304553/2010-7.\end{acknowledgments}
%
%

%
%
\end{document}